\documentclass{emulateapj}

\usepackage{float}
\usepackage{amssymb}
\usepackage{color}
\usepackage{ulem}
\usepackage{amsmath}
\usepackage{units}
\usepackage{natbib}
\usepackage{hyperref} 
\usepackage{url}
\usepackage{amstext}

\shorttitle{Bow shock fragmentation driven by a thermal instability}
\shortauthors{Suzuki-Vidal et al.}

\begin{document}


\title{Bow shock fragmentation driven by a thermal instability in laboratory-astrophysics experiments}



\author{F. Suzuki-Vidal$^{1}$, S. V. Lebedev$^{1}$, A. Ciardi$^{2,3}$, L. A. Pickworth$^{1,}$\altaffilmark{a}, R. Rodriguez$^{4}$, J. M. Gil$^{4}$, G. Espinosa$^{4}$, P. Hartigan$^{5}$, G. F. Swadling$^{1,}$\altaffilmark{a}, J. Skidmore$^{1,}$\altaffilmark{b},  G. N. Hall$^{1,}$\altaffilmark{a}, M.~Bennett$^{1}$, S.~N.~ Bland$^{1}$, G.~Burdiak$^{1}$, P.~de~Grouchy$^{1}$,\altaffilmark{c}, J. Music$^{1,}$, L. Suttle$^{1}$, E. Hansen$^{6}$ and A. Frank$^{6}$}

\affil{$^{1}$Blackett Laboratory, Imperial College London, Prince Consort Road, London SW7 2BW, United Kingdom \\ 
f.suzuki@imperial.ac.uk  \\
$^{2}$Sorbonne Universit{\'e}s, UPMC Univ. Paris 6, UMR 8112, LERMA, F-75005, Paris, France \\
$^{3}$LERMA, Observatoire de Paris, PSL Research University, CNRS, UMR 8112, F-75014, Paris France \\
$^{4}$Departamento de Fisica de la Universidad de Las Palmas de Gran Canaria, 35017 Las Palmas de Gran Canaria, Spain \\
$^{5}$Department of Physics and Astronomy, Rice University, 6100 S. Main, Houston, TX 77521-1892, USA \\
$^{6}$Department of Physics and Astronomy, University of Rochester, Rochester NY 14627, USA}


\altaffiltext{A}{Present address: Lawrence Livermore National Laboratories, CA, US.}
\altaffiltext{B}{Present address: AWE Aldermaston, UK.}
\altaffiltext{C}{Present address: Cornell University, NY, US.}

\begin{abstract}

The role of radiative cooling during the evolution of a bow shock was studied in laboratory-astrophysics experiments that are scalable to bow shocks present in jets from young stellar objects. The laboratory bow shock is formed during the collision of two counter-streaming, supersonic plasma jets produced by an opposing pair of radial foil Z-pinches driven by the current pulse from the MAGPIE pulsed-power generator. The jets have different flow velocities in the laboratory frame and the experiments are driven over many times the characteristic cooling time-scale. The initially smooth bow shock rapidly develops small-scale non-uniformities over temporal and spatial scales that are consistent with a thermal instability triggered by strong radiative cooling in the shock. The growth of these perturbations eventually results in a global fragmentation of the bow shock front. The formation of a thermal instability is supported by analysis of the plasma cooling function calculated for the experimental conditions with the radiative packages ABAKO/RAPCAL.

\end{abstract}


\section{Introduction}

One of the characteristic features of protostellar jets is the presence of shocks. They can be seen as large-scale terminal bow shocks or \textit{working surfaces} (known as Herbig-Haro or HH objects) that form as the jet interacts with previous jet ejections. Smaller-scale, internal shocks are also present which are driven by highly-variable flow velocity in the jet. These are formed as the flow moves away from the protostar and encounters and overtakes slower material from previous ejections, with the process repeating along the jet beam. Proper motion measurements of the jets in HH 34 \citep{Reipurth2002} and HH 111 \citep{Hartigan2001} show the jet flow can reach peak velocities of $\sim$200$-$300 km~s$^{-1}$ with a typical velocity variation along the jet of $\sim$40 km~s$^{-1}$.

Shocks from protostellar jets exhibit complex dynamics in which different effects such as shear, hydrodynamic instabilities and radiative cooling can be present simultaneously \citep{Hartigan2003}. In this work we are particularly interested in the effect of radiative cooling, as it can drastically modify the shock morphology. Radiative losses are strongly dependent on the opacity \citep{Drake2006book}, and if the shock region is optically thin, then radiation can escape the shock leading to an increase in the post-shock density. As the shock cools down it can be prone to the growth of thermal instabilities \citep{Field1965, Hunter1970} which can fragment and ultimately break up the shock. This effect has previously been studied mostly through numerical simulations (see e.g.~\cite{Blondin1989, Blondin1990}, \cite{Stone1993}, \cite{deGouveiadalPino1993}, \cite{Frank1998}, \cite{Tesileanu2008} and \cite{Asahina2014}). 

In this paper we describe laboratory experiments which provide a complementary approach to study the effects of radiative cooling on the structure of the bow shocks. The similarity in the key dimensionless parameters characterizing our experiments, such as the Mach number and the cooling parameter, mean that the results are scalable to the internal shocks observed in YSO jets. We observe the formation of a bow shock in the flow and its subsequent fragmentation, consistent with the onset of thermal instabilities. The bow shock in the experiments is formed from the interaction of two counter-streaming flows, which is equivalent to observations of internal shocks in YSO jets from a reference frame moving with the shock. An overall approach to modelling astrophysical phenomena in laboratory experiments is presented in the review by \cite{Remington2006}, while a recent review of the synergy between observations, theory and experiments relevant to the studies of YSO jets can be found in \cite{Frank2014}.

\begin{figure}[h!]
\epsscale{1.1}
\plotone{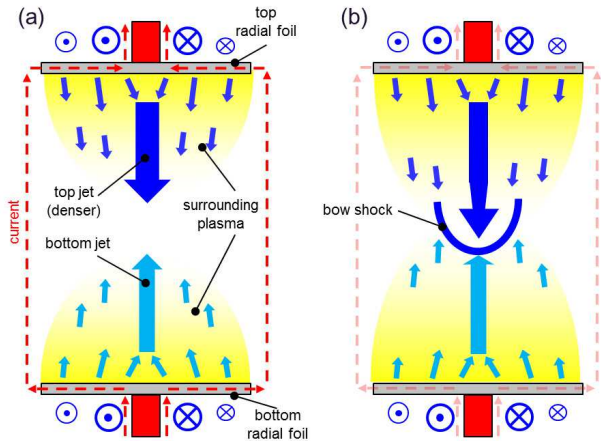}
\caption{Schematic experimental configuration to study the formation of a bow shock from the interaction between two counter-streaming jets with different relative axial velocities, represented as opposite vertical arrows on axis. The schematic depicts a side-on (radial), cut view of the system which has azimuthal symmetry. The dashed (red) arrows represent the path of the current that drives the two plasma flows. The (blue) arrows pointing into and out of the page correspond to the azimuthal magnetic field generated by the current, which provides the driving force for the two outflows. The jets are surrounded into lower density plasma (yellow regions) which moves with the same axial velocity as the jets. Smaller arrows in these regions represent the plasma flow direction. The images depict the two counter-streaming outflows (a) before their collision, (b) after they collide, triggering the formation of a bow shock moving towards the bottom foil.
\label{Fig1}}
\end{figure}

\section{Experimental setup}

The experimental setup is represented schematically in Fig.~\ref{Fig1}. Two counter-streaming, supersonic plasma outflows are produced using two co-axial and oppositely facing \textit{radial foil Z-pinches} \citep{SuzukiVidal2009apss}. Each radial foil is a metallic disc (40 mm diameter, 14 $\mu$m thick aluminum), subjected to a fast-rising electrical current pulse from the MAGPIE generator \citep{Mitchell1996}, which for the present experiments used a peak current of $\sim$1 MA in $\sim$330 ns. The current is driven to the foils through 6.35 mm diameter stainless steel tubes touching each foil at its centre, with the same current going through both of the foils via vertical posts (shown schematically in Fig.~\ref{Fig1}a). The distance between the two foils surfaces was $\sim$30 mm.

The plasma is produced by continuous ablation of the surfaces of the foils as they are heated by the current, and the ablated plasma is accelerated by the axial pressure gradient produced by the current-induced azimuthal magnetic field as it diffuses through the foils. Each of the foils produces a supersonic plasma outflow which consists of a dense central jet surrounded by lower-density ambient plasma. Both components propagate with the same axial velocity of $\sim$50$-$100 km~s$^{-1}$. Details on the formation of the outflows by a single radial foil Z-pinch can be found in \cite{SuzukiVidal2009apss}, \cite{ Ciardi2009apjl}, \cite{Gourdain2010pop} and \cite{SuzukiVidal2012pop}. In the experiments presented here, the interaction of the jets was diagnosed from the side-on (radial) direction using an optical framing camera (Invisible Vision UHSi 12/24) that imaged the optical self-emission from the plasma. This camera is capable of taking up to 12 images per experiment, with 5 ns exposure and 30 ns inter-frame separation. We also used simultaneous optical laser shadowgraphy and interferometry ($\lambda$=532 nm, pulse duration 0.3 ns). The latter diagnostic was used to measure the electron density distribution of the different plasma features present, i.e.~jets, surrounding plasma and bow shock \citep{Swadling2014rsi}.

\begin{figure*}
\epsscale{0.9}
\plotone{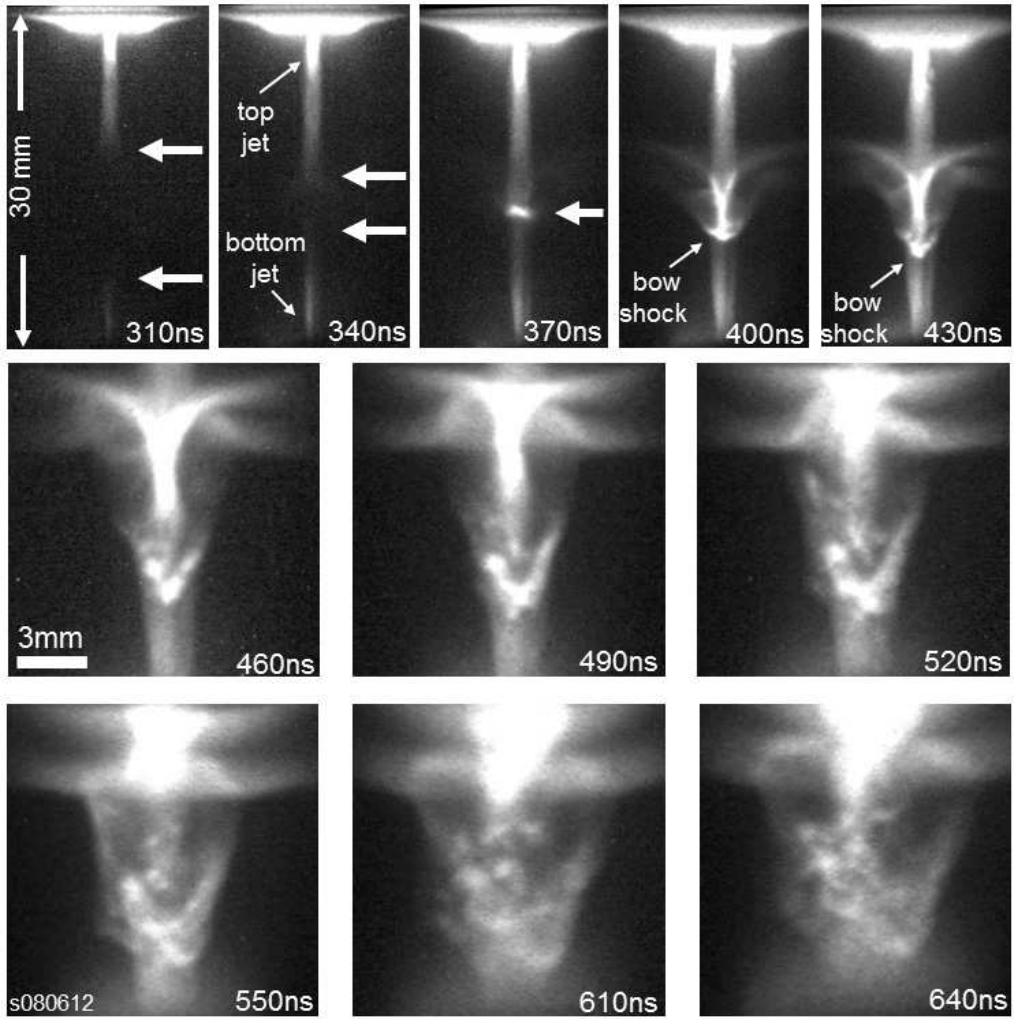}
\caption{Counter-streaming jet interaction results from optical self-emission of the plasma obtained from the same experiment. The arrows in the first three frames indicate the position of the tip of both jets (visibility dependant on image contrast levels), with their collision highlighted at 370 ns. The last six frames are focused on the bow shock region, which is seen to fragment most evidently in the last three times.
\label{Fig2}}
\end{figure*}

\subsection{Scaling}
\label{sec_scaling}

Jets from young stellar objects are generally well described by ideal magneto-hydrodynamics (MHD), and the experiments presented here are designed to produce flows in a similar regime. It is evident that these evolve over hugely different length and time-scales, and have different physical characteristics in terms of density, temperature and chemical composition. Nevertheless, the invariance properties of the ideal MHD equations provide a framework that allows meaningful scaling of the jet dynamics over many orders of magnitude (see e.g.~\cite{Ryutov1999apj, Ryutov2000apjs, Ryutov2001pop}), provided that certain constraints between the flow variables are satisfied. It is important to note that such invariance is also applicable to flows with shocks and to radiative flows under a more restrictive set of constraints \citep{Falize2011apj}.

In order for a fluid description to be applicable, we require the flows to have a \textit{localisation parameter} $\delta\ll1$. This is equivalent to an ion mean free path much less than the characteristic spatial scale of the system. Using the parameters from Table 1 in \cite{Hartigan2009apj} we estimate jets from YSOs have $\delta\sim10^{-5}$. Applicability of an MHD description also requires that the transport of momentum, magnetic field and thermal energy occurs predominantly through advection with the flow, i.e.~negligible dissipation through viscosity (Reynolds number $Re\gg1$), magnetic diffusivity (magnetic Reynolds number $Re_M\gg1$) and heat conduction (Peclet number $Pe\gg1$) respectively.  Due to their large spatial scales, YSO jets are characterised by $Re, Re_M, Pe\gtrsim10^{7}$. Our experiments are characterised by $\delta\sim10^{-4}$, $Re\sim10^5$, $Pe\sim10^3$ and $Re_M\sim10^3$ \citep{SuzukiVidal2012pop} and thus we expect a similar overall physical behaviour. 

Additionally, the interaction of radiative jets with the interstellar medium can be broadly classified by three dimensionless parameters \citep{Blondin1990}: the Mach number $M$ (the ratio of flow speed to sound speed), the density contrast $\eta$ (the ratio of density between the flow and the ambient where it propagates), and the cooling parameter $\chi_{cool}$ (the ratio of the cooling time $\tau_{cool}$ to the characteristic hydrodynamical time of the flow $\tau_{hydro}$), which quantifies the effect of radiative losses in the plasma. The specific values of $M, \eta$ and $\chi_{cool}$ essentially determine the overall morphology of the flow. Internal bow shocks in YSO jets are characterised by a relative velocity of $\sim$40 km~s$^{-1}$ and $M\sim10$, and we can expect $\eta\sim1$ and $\chi_{cool}\lesssim1$  \citep{Hartigan2009apj}. As will be discussed later in the paper, the values of these three parameters in our experiments are close to those in YSO jets.

Overall, the similarity of the dimensionless parameters allows applying the Eulerian scaling relations described in detail in \cite{Ryutov1999apj, Ryutov2000apjs, Ryutov2001pop}. Flows with identical Mach numbers will evolve with identical morphology, but on different hydrodynamic temporal ($\tau_{hydro}$) and spatial ($r_{jet}$) scales, related via the corresponding flow velocities ($V_{flow}$) as $\tau_{hydro}=r_{jet}/V_{flow}$. Taking the jet radius as a characteristic spatial scale, young stellar jets typically have $r_{YSO}\sim50$ AU $\sim 10^{15}$~mm, and experiments $r_{exp}\sim1.5$~mm), flow velocities of $V_{YSO}=40$ km~s$^{-1}$ (i.e.~typical velocity variability in YSO jets) and for the experiments here $V_{exp}=140$ km~s$^{-1}$ (i.e.~the relative velocity of a single jet in our experiments, see explanation in the next section). Thus the characteristic temporal scales are $\tau_{YSO}=$1.9$\times10^8$ s ($\sim$6 years) and $\tau_{exp}=10$ ns respectively. The total time interval over which the evolution of the flows is followed in the experiments of $\sim$300 ns ($\approx30~\tau_{exp}$) thus corresponds to $\sim$180 years of evolution for the astrophysical counterpart, which exceeds the typical time scale for multi-epoch observations of YSOs with the Hubble Space Telescope of $\sim$10 years (see e.g.~\cite{Hartigan2011apj}).

\section{Experimental results and discussion}

The overall evolution of the interaction of the two counter-streaming plasma jets can be seen in Fig.~\ref{Fig2}, which shows side-on optical emission images obtained in the same experiment at different times after the start of the current pulse driving the jets. Images in the top row of Fig.~\ref{Fig2} show the entire region between the two foils and correspond to the early times of the interaction (310$-$430 ns). The images show the formation of two well-collimated jets on the axis of the system propagating towards each other. Although the top and bottom foils are nominally the same and are driven by the same current, an asymmetry between the top and bottom flows is evident from the emission images. The top jet is formed first compared to the bottom jet, which is seen by looking at the positions of the tips of the jets indicated by horizontal arrows on the first 3 frames. This asymmetry is caused by opposite polarity current drive for each foil (radially outwards on the bottom foil compared to radially inwards on the top foil) and is reproducible from experiment-to-experiment. The asymmetry affects the details of the plasma ablation and initial acceleration of the flow at the foil \citep{Gourdain2013prl}. This results in different flow parameters (e.g.~ram pressures) which, as the jet collide, lead to the formation of a bow shock. This is first evidenced at 370 ns as a highly-emitting region aligned with the head-on collision between the two jet tips. The bow shock is fully formed at 400 ns and is seen to move downwards, i.e.~towards the bottom foil. At this time the post-shock region is seen as a highly-compressed, highly-collimated column, while the bow shock begins to develop small-scale structures from here onwards. The long-term evolution of the bow shock and of these structures is highlighted in the subsequent images, corresponding to times 460$-$640 ns.

\begin{figure*}
\epsscale{1.1}
\plotone{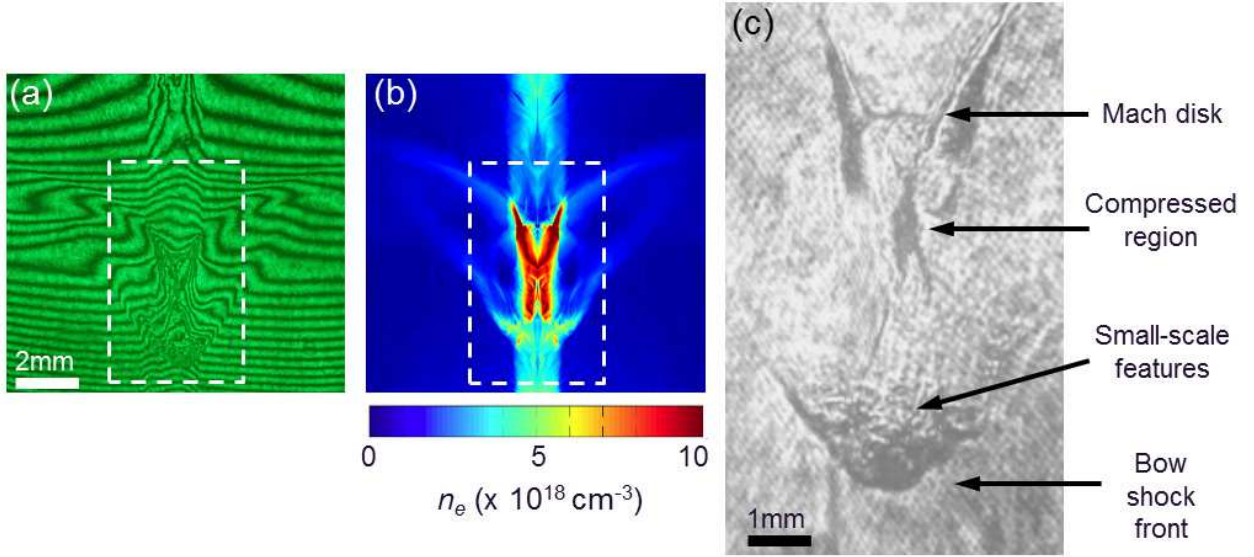}
\caption{Optical laser probing of the bow shock at 400 ns (same experiment as Fig.~\ref{Fig2}). (a) Raw data from laser interferometry, (b) axi-symmetric electron density ($n_e$) from analysis of (a). (c) Laser shadowgraphy, with the field of view shown schematically in the dashed inset in (a) and (b). The fringing on this image is an artefact present in this particular diagnostic.
\label{Fig3}}
\end{figure*}

The first collision between the two jets seen in the image at 370 ns occurs $\sim$2 mm below the mid-plane, which suggests that the two jets have slightly different velocities. This is confirmed by following the positions of the visible tips of the jets in time  before their collision, resulting in tip velocities of $V_{top~jet}\sim80\pm10$~km~s$^{-1}$ and $V_{bottom~jet}\sim60\pm10$~km~s$^{-1}$. The higher velocity in the top jet, combined with a larger density inferred from the stronger self-emission and from laser interferometry, imply an imbalanced ram pressure ($P_{ram}=\rho V^2$) in the collision, which explains the shape and orientation of the bow shock and its downward propagation. The velocity of the leading edge of the bow shock was measured as $V_{bow}\sim40\pm10$~km~s$^{-1}$.

The velocity of the bow shock measured in the experiments can be compared with the velocity expected from a one-dimensional, momentum flux conservation argument (\cite{Norman1983NWS, Hartigan1989apj, Blondin1990, deGouveiadalPino1994, deGouveiadalPino2005, Nicolai2008}). This is convenient to do in the reference frame of a single jet driving a shock through a stationary medium ahead of it. The velocity of a bow shock $V_{bow}$ is related to the velocity of the jet $V_{jet}$ and the density contrast $\eta$ between the jet and the pre-shock external medium by $V_{bow}\approx V_{jet}$(1+$\eta^{-1/2}$)$^{-1}$. The measured experimental jet tip velocities of $V_{jet~top}\sim$80 km~s$^{-1}$ and $V_{jet~bottom}\sim$60 km~s$^{-1}$ for the top and bottom jets respectively can be translated into a single jet with a relative tip velocity of V$_{jet~rel}$=140~km~s$^{-1}$ which in turns forms a bow shock with a relative velocity of $V_{bow~rel}$=100~km~s$^{-1}$. Using these relatives velocities in the equation above result in a density contrast $\eta\sim6$, i.e.~the experiment is equivalent to the interaction of a single jet which is 6 times denser than the ambient medium where the shock is generated which, for this particular counter-streaming geometry, the ambient medium is the opposite (bottom) jet. We can calculate the internal Mach number of the jet working surface by taking the ratio of the jet relative velocity (V$_{jet~rel}$=140~km~s$^{-1}$) and the typical ion acoustic speed in the jet flow of $c_s\sim$15 km~s$^{-1}$ \citep{SuzukiVidal2012pop}, resulting in an internal Mach number of $M\sim10$. Therefore the values of $\eta$ and $M$ in the experiments are similar to those in YSO jets \citep{Hartigan2009apj} and thus the experiments should evolve with a similar jet/shock morphology as discussed earlier.

In addition to the bow shock, the images in Fig.~\ref{Fig2} also show the collision between the two flows off-axis, i.e.~the plasma surrounding the jets, which leads to the formation of a \textit{double-shock} structure that extends at large radii. This feature is especially evident at the top of the images from $\sim$460 ns onwards. These standing shocks remain approximately symmetric with respect to the mid-plane between the two foils. Preliminary MHD simulations indicate that these features arise due to the presence of toroidal magnetic field advected by the counter-streaming flows \citep{SuzukiVidal2014aps}. A more detailed discussion of the pile up of the magnetic fields will be published separately.

\subsection{Bow shock and jet working surface: evolution and small-scale structures}
 
The emission images in Fig.~\ref{Fig2}, show in detail the evolution of the bow shock from its formation at 400 ns through its rapid fragmentation. At 400 ns the surface of the bow shock and the post-shock region just behind it are smooth. The formation of small-scale structures occurs on timescales that are shorter than the inter-frame separation for this particular diagnostic. This gives an upper limit for their development timescale of $\sim$30 ns, which is consistent with the appearance of new structures between frames at later times. The characteristic spatial scale of the structures seen in the images ranges from $\sim$200 $\mu$m to $\sim$2 mm, and at the final stages of the evolution (the last two panels at 610 ns and 640 ns) the bow shock has fragmented into a large number of emitting \textit{clumps}. The smallest observed size ($\sim$200 $\mu$m) seen on the images obtained with the optical framing camera is comparable with the spatial resolution of this diagnostic, and is also comparable with the motional blurring due to the temporal resolution of the diagnostic (5 ns $\times$ 40 km~s$^{-1}$ = 200 $\mu$m).
 
Further details of the small-scale structures present in the bow shock and post-shock region were obtained with laser probing (0.3 ns pulse duration), which has significantly better spatial resolution ($\lesssim50~\mu$m) and reduced motional blurring ($\sim10~\mu$m). The interferometry channel of this diagnostic provides measurements of the spatial distribution of the plasma electron density, and the shadowgraphy channel, which is sensitive to spatial gradients of the electron density, provides information on the characteristic spatial scales of the non-uniformities. Fig.~\ref{Fig3} shows laser probing images obtained at 400 ns in the same experiment as the optical emission images shown in Fig.~\ref{Fig2}, i.e.~the timing of the laser probing images corresponds to the fourth panel in Fig.~\ref{Fig2}. Fig.~\ref{Fig3}a shows the raw image from the interferometry channel. The apparent distortion of the interference fringes, which were initially horizontal and uniformly spaced, is caused by changes in the interference state which are induced by a spatially varying phase delay imparted on the probe beam by the plasma. This phase delay can be extracted from the image \citep{Swadling2014rsi} and is proportional to the electron column density of the plasma, $n_eL$ (in cm$^{-2}$). Here $L$ is the length of plasma along the probing beam, which changes as a function of position in the plane of the image. The sufficiently good axial symmetry of the object allows applying Abel inversion \citep{Hutchinson2005book} to the electron column density, resulting in the axi-symmetric (radial) distribution of electron density $n_e(r)$ (in cm$^{-3}$) shown in Fig.~\ref{Fig3}b. The highest electron density near the axis, reaching $n_e\sim$10$^{19}$ cm$^{-3}$ (up to 75\% uncertainty due to errors in the choice of the central axis and left-right asymmetries), is observed in the narrow, compressed region formed above the bow shock (the post-shock region), while the electron density in the flow below the bow shock (the pre-shock region) is $n_e\sim$(3$\pm$2)$\times$10$^{18}$ cm$^{-3}$ (i.e.~$\sim$60\% uncertainty). We note that, although these errors seem large, for measurements of electron density off-axis these errors decrease rapidly and can reach values $\lesssim$20\%. The higher electron density present in the jet driving the bow shock is qualitatively consistent with the density contrast (heavy jet propagating through ambient) inferred from application of the one-dimensional momentum flux conservation argument introduced in the previous section.

The laser shadowgraphy diagnostic in Fig.~\ref{Fig3}c shows several interesting features. Firstly, there are two dark regions in the top part of the image positioned on both sides from the axis, and just below them a dark vertical region on axis. These features are produced by large density gradients and their shape is consistent with compression of the jet to a diameter smaller than the initial jet diameter due to converging flows driven by large post-shock pressures. Formation of such shocks was observed in numerical simulations of high Mach number jets propagating through ambient media (e.g.~\cite{Norman1982}), and the structure we see in the experiment is very similar to that observed in simulations of radiatively cooled jets with a similar density contrast, e.g.~presented in Fig.~7c of \cite{Blondin1990} for a density contrast of $\eta$=3, which is of the same order to the one inferred for these experiments of $\eta$=6. The simulations by \cite{Blondin1990} also show the presence of a second standing conical shock, positioned between the bow shock and the region of maximum compression which opens up towards the bow shock. The image in Fig.~\ref{Fig3}c indicates that this conical shock is also present in the experiment, though it is less pronounced indicating a smaller density gradient than that in the downstream conical shock. We also note the presence of a horizontal dark region connecting the dark features at the top of the image, which is consistent with the expected shape of a \textit{Mach disk} (see e.g.~Fig.~1a in \cite{Hartigan1989apj} and Fig.~7b in \cite{Blondin1990}). 
	
Finally, the image in Fig.~\ref{Fig3}c shows the presence of small-scale structures already developing behind the bow shock at this early stage of the evolution. The perturbations appear to be elongated, predominantly in the direction normal to the jet axis, and the smallest detectable spatial scale is $\sim$120$-$170 $\mu$m, which is a factor of $\sim$3 larger than the spatial resolution of the diagnostic ($\sim$50 $\mu$m in the diffraction limited case). We interpret the observed development of small-scale perturbations in the bow shock leading to its complete fragmentation later in time as being a result of strong radiative cooling in the post-shock plasma. The fragmentation of the bow shock and the development of dense clumps in this region are associated with a thermal instability \citep{Field1965}, which is discussed thoroughly in the next section. The fragmentation of shocks in protostellar jets due to thermal instabilities has been previously studied by numerical simulations (see e.g.~\cite{Blondin1989, Blondin1990, Frank1998}), however the effect of this instability in YSO jets and shocks is still unclear.

\subsection{Radiative cooling and thermal instabilities in the bow shock}

The most striking result in the experiments is the rapid development of small-scale spatial structures in the bow shock and post-shock region. Our interpretation is that the fragmentation of the bow shock is related to a dynamic, local thermal instability that leads to the condensation of density perturbations by radiative cooling (see e.g.~\cite{Field1965}, \cite{Mathews1978}, \cite{Fall1985}, \cite{Balbus1986} and \cite{Blondin1989densitypert}). The instability develops over the characteristic radiative cooling time, $\tau_{cool}$, which for optically thin plasmas, such as the ones in our experiments \citep{Espinosa2014hedp}, is given by the ratio of thermal energy density, $U$, to the radiated power per unit volume $P_{rad}=n_{e}n_{i}\Lambda(n_i,T_{e})$ as $\tau_{cool}=U/n_{e}n_{i}\Lambda(n_i,T_{e})$, where $\Lambda(n_i,T_{e})$ is the normalised cooling function (in erg cm\textsuperscript{3} s\textsuperscript{-1}), and $n_i$ is the ion density.  This expression can also be written as \citep{Ryutov1999apj}:

\begin{equation}
\tau_{cool}[s]=2.4\times10^{-12}\frac{(\bar{Z}+1)T[eV]}{\bar{Z}n_{i}[cm^{-3}]\Lambda(n_i,T_{e})}
\label{eq_tcool}
\end{equation}

\noindent where $\bar{Z}$ is the average ionization in the plasma.

\begin{figure*}
\epsscale{1.1}
\plotone{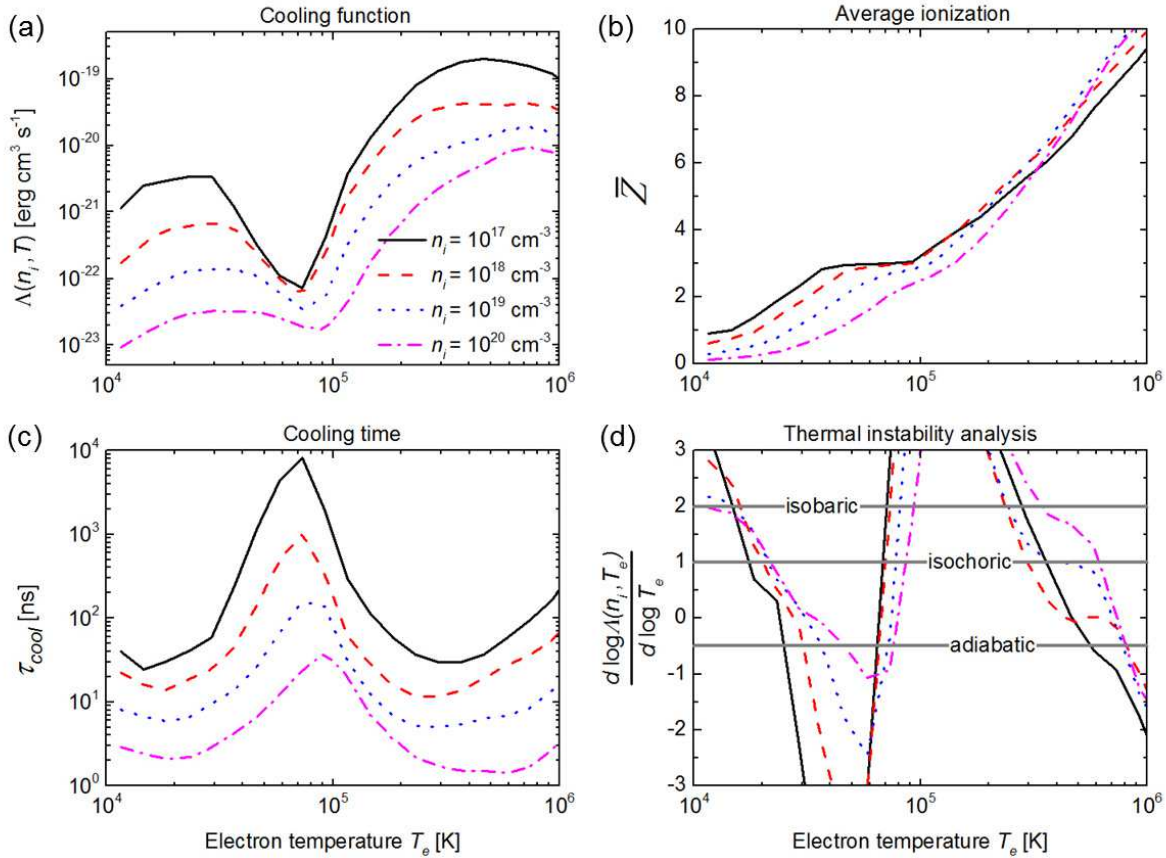}
\caption{Calculated (a) cooling function and (b) average ionization for aluminum as a function of electron temperature for different ion densities relevant to our experiments (the legend shown in (a) applies to the entire figure). (c) Cooling time calculated using Eq.~\ref{eq_tcool}. (d) Thermal instability analysis applied to the cooling functions shown in (a). The regions below each horizontal line indicate the thresholds for the onset of isobaric ($<2$), isochoric ($<1$) and adiabatic ($<-0.5$) thermal instabilities.}
\label{Fig4}
\end{figure*}

When radiative cooling occurs on time scales that are long compared to the sound speed crossing time, perturbations over a region of size $\lambda_{iso}=c_{s}\tau_{cool}$ (where $c_{s}$ is the ion acoustic speed) tend to maintain a common pressure (isobaric). In this case $nT\sim$constant and the cooling rate then scales as $\Gamma=\tau_{cool}^{-1}\propto\Lambda(T)/T^{2}$. In the isobaric regime, if the cooling rate increases for a decreasing temperature, i.e. $d\Gamma/dT<0$, then radiation losses will be even more efficient in removing energy from the plasma and further reducing its temperature. To maintain pressure equilibrium with its surroundings the density increases, thus further increasing the radiated power losses and potentially leading to a run-away condensation of the initial density perturbations. Detailed analysis of the instability was done by \cite{Field1965}, \cite{Hunter1970} and \cite{Balbus1986}. For isobaric modes, the condition for stability is given by

\begin{equation}
\frac{d( log\Lambda(n_i,T))}{d(logT)}<2
\label{eq_stability}
\end{equation}

While $\lambda_{iso}$ sets an upper limit for the the length-scale of isobaric condensation, thermal conduction will suppress short wavelength perturbations that are of the order of the so-called Field's length ($\lambda_{Field}=2\pi[(\gamma-1)\chi_{th}\tau_{cool}]^{1/2}$), where $\chi_{th}=\kappa/n_{e}$ is the thermal diffusivity and $\gamma$ is the ratio of heat capacities. The most unstable wavelength, valid in the regime of large wave-numbers (but still smaller than a critical wavenumber $k_{crit}=2\pi\lambda_{Field}^{-1}$), is given by the geometric mean \citep{Field1965} $\lambda_{max}=(\lambda_{Field}\lambda_{iso})^{1/2}$ so that density and temperature perturbations will be unstable in the wavelength range 

\begin{equation}
\lambda_{Field}<\lambda<\lambda_{iso}
\label{eq:wavelength condition}
\end{equation}

Although cooling in our experiments is different from that in astrophysics due to the differences in elements and physical conditions of temperature and densities, for optically thin plasmas the relevant parameters for comparison are the dimensionless cooling parameter $\chi_{cool}$ and the dependence of the cooling function $\Lambda(n_i,T)$ on temperature. To estimate the effects of radiative cooling in the post-shock aluminum plasma, we use new cooling rates calculated with the computational packages ABAKO/RAPCAL \citep{Rodriguez2008lpb, Florido2009pre}. The codes calculate the plasma level populations and average ionizations by solving the set of rate equations of the collisional-radiative model implemented in ABAKO (assuming the plasma is optically thin and in steady-state). The model is capable of accounting for coronal equilibrium, local and non-local thermodynamic equilibrium regimes. The atomic data required were obtained using the FAC code \citep{Gu2008canada} in the relativistic detailed configuration accounting approach, with the spin-orbit split arrays formalism \citep{BaucheArnoult1985} and including configuration interaction within the same non-relativistic atomic configurations. The databases of cooling functions and average ionizations obtained with ABAKO/RAPCAL were subsequently parametrized as a function of the plasma density and temperature using the PARPRA code \citep{Rodriguez2014comm}. Results from these numerical calculations are presented in Figs.~\ref{Fig4}a-b showing the variation of the cooling function $\Lambda(n_i,T_{e})$ and the average ionization $\bar{Z}$ for an aluminum plasma as function of electron temperature (ranging from $\sim10^{4}-10^{6}$ K, $\sim1-85$ eV) for different ion densities typical of our experiments ($n_i=10^{17}-10^{20}$ cm$^{-3}$). Fig.~\ref{Fig4}a shows the rate of cooling increases with decreasing ion density, varying up to two orders of magnitude. The average ionization in Fig.~\ref{Fig4}b shows an increase with temperature, with little variation as a function of ion density. With the cooling function and the average ionization we can calculate the cooling time $\tau_{cool}$ (Eq.~$\ref{eq_tcool}$), and this is plotted for different aluminum ion densities in Fig.~\ref{Fig4}c. The plot shows that overall the cooling time decreases with increasing ion density, as $\tau_{cool}\propto1/\Lambda(n_i,T_e)$, and longer cooling times can be expected around an electron temperature of $T_e\sim$10$^5$~K. With the cooling time we can calculate the cooling parameter $\chi_{cool}=\tau_{cool}/\tau_{hydro}$ which quantifies the importance of radiative losses in the plasma. Taking the estimated hydrodynamical time of the experiments of $\tau_{exp}\sim10$ ns (see Sec.~\ref{sec_scaling}), we see that, independent of ion density $\tau_{cool}>\tau_{exp}$ in the region around $T_e\sim$10$^5$~K, which implies $\chi_{cool}\gtrsim1$ and thus radiative losses are not important. This region correlates with a dip in the cooling functions at this temperature. Besides this temperature region, however, the cooling time is overall of the order of the hydrodynamical time and thus the cooling parameter $\chi_{cool}\sim1$. The hydrodynamical time can also be estimated as the time it takes for the shock to form and become fully disrupted, which from Fig.~\ref{Fig2} it can be taken as $\tau_{exp}\sim550-370~ns=180~ns$ resulting in $\chi_{cool}\sim(3-30~ns)/180~ns<1$ and thus the plasma in the shock is expected to be radiatively cooled.

Knowledge of the dependence of the cooling function as a function of temperature allows performing a thermal instability analysis. The stability condition for isobaric modes presented in Eq.~\ref{eq_stability} ($dlog\Lambda(n_i,T_e)/dlogT_e<2$) is plotted in Fig.~\ref{Fig4}d using the cooling functions for different ion densities in Fig.~\ref{Fig4}a. The threshold for isobaric instabilities shows that, almost independently of the ion density, the plasma is expected to become unstable at temperatures of $T_{e}\sim(1.5-9)\times10^{4}\,\unit{K}$ ($\sim5$ eV) and at $T_{e}\gtrsim2.5\times10^{5}\,\unit{K}$ ($\gtrsim20$ eV). Following the analysis in \cite{Shchekinov1978} we also plot other instability thresholds, namely isochoric and adiabatic modes, which correspond to $dlog\Lambda(n_i,T_e)/dlogT_e<1$ and $<-0.5$ respectively. 

Furthermore, the onset of thermal instabilities in the post-shock plasma depends on the thermal evolution of the ion and electron plasma components. Because of the large difference between the ion and electron masses, there is a large temperature difference between the ion and electron components immediately behind the bow shock. In the strong shock approximation (Mach number $M\gg1$), the post-shock (PS) ion temperature is $T_{i,PS}=2(\gamma-1)Am_{p}V_{bow~rel}^{2}/k_{B}(\gamma+1)^{2}\sim10^{6}\,\unit{K}$, where $A$ is the atomic weight ($A$=27 for aluminum), $m_{p}$ is the proton mass, $V_{bow~rel}$ is the shock velocity in the reference frame of a stationary pre-shock medium (i.e.~$V_{bow~rel}=100$~km s$^{-1}$), and $\gamma=5/3$ assuming an ideal gas. Compared to the ions, the electrons are compressed adiabatically across the shock and their temperature increases only by a factor of $4^{\gamma-1}\sim2.5$ to $T_{e,PS}\sim(2-3)\times10^{5}\,\unit{K}$. Although the plasma can be unstable at these temperatures, the existence of a well-defined unstable range of wavelengths, as given in Eq. \ref{eq:wavelength condition}, is not satisfied. The post-shock is characterized by a relaxation region where the energy exchange between ions and electrons competes with radiative cooling. Immediately after the shock the electrons are thermally decoupled from the ions and their cooling time-scale ($\lesssim1\,\unit{ns}$) is much shorter than the ion-electron energy equilibration time-scale ($\tau_{\epsilon}^{i/e}=\tau_{\epsilon}^{e/i}/\overline{Z}\sim20\,\unit{ns}$). Thus, over time-scales of a nanosecond, the electron temperature rapidly drops while the ion temperature remains essentially constant. The cooling rate also decreases considerably and an equilibrium is immediately reached behind the shock where electron heating due to the energy received by the ions balances their radiative energy losses 

\begin{equation}
\frac{T_{i}-T_{e}}{\tau_{\epsilon}^{e/i}}\approx(\gamma-1)\frac{n_{i}}{k_{B}}\Lambda\left(n_i,T_e\right)
\label{eq_TiTe}
\end{equation}

In this regime $\tau_{\epsilon}^{i/e}\approx\tau_{cool}\sim15-20\,\unit{ns}$ and the electrons are approximately isothermal. Detailed calculations of temperature equilibration show that the equilibrium electron temperature in the post-shock relaxation layer is $T_{e}\sim(1-1.5)\times10^{5}\,\unit{K}$, thus the plasma is expected to be thermally stable. After $\sim40\,\unit{ns}$ the electron and ion plasma components are fully equilibrated and their common temperature decreases below $\lesssim7\times10^{4}$ K. This places the plasma in conditions corresponding to a thermally unstable regime. The condition on the wavelengths is satisfied and density and temperature perturbations with wavelengths $\lambda\sim30-100\,\unit{\mu m}$ grow over very short time-scales of order $\lesssim1$ ns.

\begin{figure*}
\epsscale{1.1}
\plotone{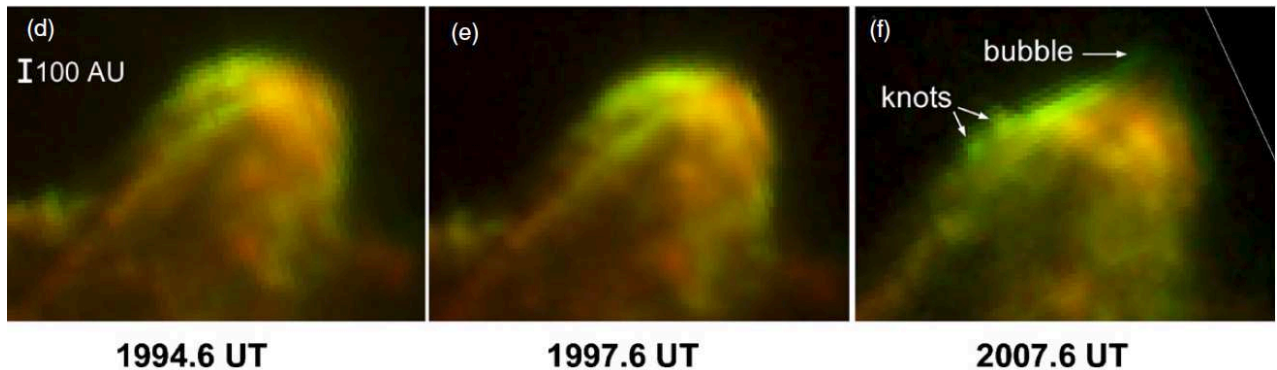}
\caption{Multi-epoch observations of HH~1 showing the formation of knots in the bow shock (adapted from Fig.~2 in \citep{Hartigan2011apj}). The image shows two out of three epochs of HST images, with H$_\alpha$ in green, the $[S_{II}]$ in red, and yellow denotes emission in both filters.}
\label{Fig5}
\end{figure*}

\section{Conclusions}

We presented a new experimental configuration that aims at studying experimentally the formation of bow shocks relevant to those present in young stellar jets. In our experiments we produce two counter-propagating, supersonic plasma jets from plasma ablation of two radial foil Z-pinches. A bow shock is driven by the head-on collision between the two jets and the bow shock properties are determined by their relative velocities and densities. A key result is that initially the bow shock is smooth but then quickly, within time scales of $\sim$30 ns, develops small-scale spatial features behind the shock front. This time-scale is consistent with the expected cooling times in the experiments, which were estimated using the cooling functions and average ionization calculated with the radiative packages ABAKO/RAPCAL. This allowed performing a thermal instability analysis for isobaric modes resulting in expected temperature ranges at which the plasma should become thermally unstable. Detailed analytical calculations of the thermal evolution of the ion and electron components predict that, for the experimental shock conditions, we expect the development of an isobaric thermal instability in time scales of $\sim$40 ns, and with typical spatial scales of $\sim$30$-$100 $\mu$m. This is in very good agreement with our experimental results that show an initially smooth bow shock which, within $\sim$30 ns, fragments into small-scale spatial features with typical sizes of $\sim120~\mu$m.

Thermal instabilities may play a role in HH shocks provided the shock velocities are high enough to raise the temperature in the post-shock gas significantly above the peak of the cooling curve around $2\times 10^5$~K. The critical shock velocity for the onset of cooling instabilities has been estimated to be $\sim$ 200 km~s$^{-1}$ in 1-D simulations \citep{Smith1989}, though \cite{Sutherland2003ApJpaper2} found that thermal instabilities in their 2-D simulations generated filaments and voids in the post-shock gas for shock velocities as low as 120 km~s$^{-1}$. However, \citep{Innes1992} demonstrated that even a fairly weak pre-shock magnetic field stabilized shocks up to 175 km~s$^{-1}$. In HH jets, most shocks have velocities $\lesssim$ 100 km~s$^{-1}$, and should not be affected by thermal instabilities. However, the strongest bow shocks, such as HH~1 and HH~2, have high ionization lines \citep{Bohm1993} and broad line profiles \citep{Hartigan1987apj} indicative of shock velocities $\sim$ 200 km~s$^{-1}$, well above the criteria for thermal instabilities. As shown in Fig.~\ref{Fig5}, bright knots do appear along the bow shock in HH~1 on time-scales of decades. Using the scaling presented at the end of Sec.~\ref{sec_scaling}, the laboratory temporal scale of 30~ns would correspond to an astrophysical time scale of $\sim$ 17 yr, consistent with the astronomical observations. The spatial scales associated with the astronomical knots also scale well with the non-uniformities in the experiment. For example, the minimum spatial scale of the non-uniformities in the experiments of 0.12~mm scale to 4~AU in the astronomical observations, where the spatial resolution is $\sim$ 20~AU. The larger non-uniformities present in the optical self-emission (e.g. the last three panels of Fig.~\ref{Fig2}) have a typical size of $\sim$ 1~mm, which scales to 30~AU in Fig.~5, consistent with the size of the new knots in HH~1. Driven from the same source on the other side of the outflow, HH~2 also has a high shock velocity and shows small knots that appear and merge along the strongest shock fronts \citep{Hartigan2011apj}. Of course, there are other ways to generate clumps along bow shocks such as Kelvin-Helmholtz instabilities or even a clumpy pre-shock density, and we cannot rule out these possibilities without further observational data.

More generally, the thermal instability analysis could be applied to cooling curves used for numerical simulations of protostellar jets and shocks (see e.g.~\cite{Dalgarno1972, Kafatos1973, Sutherland1993ApJS}). Although in these cases we expect the temperature ranges for the onset of thermal instabilities to differ from those expected in the experiments due to the different elements and abundances that characterize them (e.g.~H in simulations compared to Al in the experiments), in both cases the instability should develop at the appropriate slope of the cooling curve by the condition given in Eq.~\ref{eq_stability}. 

For typical interstellar medium compositions, shock velocities of the order of $\gtrsim 120$ km~s$^{-1}$ are required to push the cooling curve into the regime where it declines rapidly enough with increasing temperature to become prone to thermal instabilities \citep{Sutherland1993ApJS, Innes1992}. Hence, most shocks in jets do not fall into this regime. However, some, like the clumps observed in HH~2, have shock velocities of at least 200 km~s$^{-1}$ \citep{Hartigan2011apj}, thus we could expect thermal instabilities to be related to the appearance of spatial inhomogeneities in the shock.

The experiments presented here are important as they shed light on the onset and non-linear evolution of the thermal instability. Numerical simulations of thermally unstable flows can be challenging as thermal conduction must be explicitly included to suppress the growth of small-scale perturbations. In particular, the Field's length has to be resolved by at least a few computational cells \citep{Koyama2004} to avoid the growth of perturbation and fragmentation at the grid scale. Furthermore, the presence of magnetic fields further increases the complexity of numerical calculations by making thermal conduction anisotropic. In that direction, similar experiments to the ones presented here can be designed to increase sufficiently the magnetic field in the flow to allow studying such regime in the future. This could be relevant to previous theoretical studies which predict the suppression of thermal instabilities by magnetic fields (e.g.~\cite{Innes1992} and \cite{Lesaffre2004_427_147}).

\acknowledgments

Work supported in part by The Royal Society through a University Research Fellowship, by EPSRC Grant No. EP/G001324/1, by DOE cooperative agreements No. DE-F03-02NA00057 and No. DE-SC-0001063, and by the LABEX Plas@Par project. Work received financial state aid managed by the Agence Nationale de la Recherche, as part of the program ``Investissements d'avenir'' under ref. ANR-11- IDEX-0004-02.

\bibliographystyle{apj}
\bibliography{MyCollection}




\end{document}